\documentclass[12pt]{iopart}

\expandafter\let\csname equation*\endcsname\relax
\expandafter\let\csname endequation*\endcsname\relax

\usepackage{amsmath}
\usepackage{bm}
\usepackage{graphicx}
\usepackage{amsfonts}
\usepackage{amssymb}
\usepackage{amsbsy}
\usepackage{array}
\usepackage{iopams}
\usepackage{dsfont}
\usepackage{subfigure}

\begin{document}

\title{Role of Energy-Level Mismatches in a Multi-Pathway Complex of Photosynthesis}

\author{James Lim$^{1,2}$, Junghee Ryu$^{1,2}$, Changhyoup Lee$^{1,2}$, Seokwon Yoo$^{1,2}$, Hyunseok Jeong$^{2,3}$ and Jinhyoung Lee$^{1,2,4}$}

\address{$^1$ Department of Physics, Hanyang University, Seoul 133-791, Korea}
\address{$^2$ Center for Macroscopic Quantum Control, Seoul National University, Seoul, 151-742, Korea}
\address{$^3$ Department of Physics and Astronomy, Seoul National University, Seoul, 151-742, Korea}
\address{$^4$ School of Computational Sciences, Korea Institute for Advanced Study, Seoul 130-722, Korea}

\ead{james83@hanyang.ac.kr and hyoung@hanyang.ac.kr}

\date{\today}

\begin{abstract}

Considering a multi-pathway structure in a light-harvesting complex of photosynthesis, we investigate the role of energy-level mismatches between antenna molecules in transferring the absorbed energy to a reaction center. We find a condition in which the antenna molecules faithfully play their roles: Their effective absorption ratios are larger than those of the receiver molecule directly coupled to the reaction center. In the absence of energy-level mismatches and dephasing noise, there arises quantum destructive interference between multiple paths that restricts the energy transfer. On the other hand, the destructive interference diminishes as asymmetrically biasing the energy-level mismatches and/or introducing quantum noise of dephasing for the antenna molecules, so that the transfer efficiency is greatly enhanced to near unity. Remarkably, the near-unity efficiency can be achieved at a wide range of asymmetric energy-level mismatches. Temporal characteristics are also optimized at the energy-level mismatches where the transfer efficiency is near unity. We discuss these effects, in particular, for the Fenna-Matthews-Olson complex.

\end{abstract}

\pacs{03.67.-a, 03.65.Yz, 87.15.hg, 05.60.Gg}

\maketitle

\newcommand{\bra}[1]{\left\langle #1\right|}
\newcommand{\ket}[1]{\left|#1\right\rangle}
\newcommand{\abs}[1]{\left|#1\right|}
\newcommand{\ave}[1]{\left<#1\right>}

\section{Introduction}

Photosynthetic complexes are sophisticated light-harvesting machinery consisting of antenna molecules. The energy absorbed by the antenna molecules, so-called exciton, is transferred to many intermediate molecules and eventually arrives at a reaction center (RC) where the process of biochemical energy conversion is initiated. Recent work has reported that quantum theory governs the exciton transfer in some light-harvesting complexes that harness the absorbed energy with almost 100{\%} efficiency~\cite{Engel2007}--\cite{Collini2010}. It has also been suggested that the interplay of quantum walk and noisy environment provides the high light-harvesting efficiency in these complexes~\cite{Mohseni2008}--\cite{Rebentrost2009}. However, the underlying mechanism of noise-assisted enhancement has remained elusive and the role of structural characteristics of antenna complexes is still not clear. First, photosynthetic complexes possess energy-level mismatches between antenna molecules. As energy-level mismatches are likely to cause Anderson localization~\cite{Anderson1958} and to inhibit the transfer of excitation~\cite{Mohseni2008}, it is desired to understand why photosynthetic complexes evolve by maintaining such energy-level mismatches instead of eliminating or reducing if impossible. Second, the noise-assisted enhancement is not possible in certain situations, such as uniform linear chains with no energy-level mismatches~\cite{Plenio2008}. These observations suggest that energy-level mismatches are related to the condition for the enhancement~\cite{Caruso2009}, but their fundamental role is still not unraveled. Moreover, even in the cases where the noise-assisted enhancement is possible, the underlying principle of enhancement remains ambiguous due to the lack of basic studies based on quantum interference~\cite{Chin2010}.

In this paper, we investigate the role of energy-level mismatches in a multi-pathway complex where multiple sub-complexes are independently connected to the RC via a receiver molecule (see \fref{fig:antenna_complexes}). For a single-pathway complex, we show that any energy-level mismatches suppress the efficiency of exciton transfer as expected by Anderson localization. In the absence of energy-level mismatches, we show that quantum noise of dephasing, which decreases the quantum coherence of excitation, will never help the transfer efficiency. For a bi-pathway complex, on the other hand, we demonstrate that the absence of energy-level mismatches leads to undesired destructive interference at the receiver molecule. The destructive interference is caused by the two types of probability amplitudes, one of coming from one pathway and the other of going to and returning from the other pathway. It blocks exciton transfer to the RC so that the transfer efficiency is less than $50{\%}$ and relaxation of the sub-complexes to the ground state becomes very slow. We show that energy-level mismatches play a dominant role in suppressing the undesired interference. Then, the presence of energy-level mismatches enhances the transfer efficiency even though the resultant localization effect suppresses the energy transfer in each pathway of the complex. Due to the competition between the Anderson localization and the destructive interference at the receiver molecule, a moderate amount of energy-level mismatches will improve the light-harvesting efficiency as minimizing both the undesired effects in the energy transfer. This picture is consistent with our results. In addition, dephasing noise is found to relaxe the localization and to suppress the undesired interference at the receiver molecule so that the cooperation of energy-level mismatches and dephasing noise significantly improves the transfer efficiency after all.

\section{Light-Harvesting complex}

\begin{figure} [t]
\centering
\includegraphics{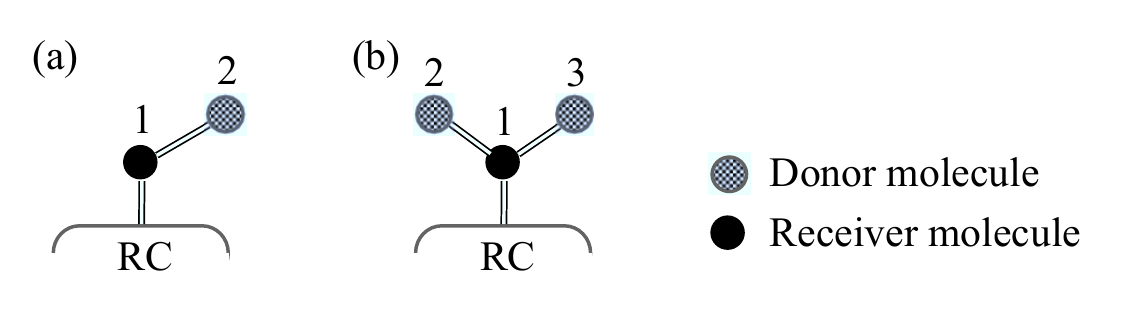}
\caption{Schematic representations of (a) the single-pathway complex and (b) bi-pathway complex. The molecule coupled with the RC is called a receiver and the other ones donors. The solid lines between molecules represent electronic coupling that enables an exciton to hop between them.}
\label{fig:antenna_complexes}
\end{figure}

Light-harvesting complex is modeled as a system consisting of $n$ two-level molecules, whose dynamics is governed by a master equation in the form of
\begin{eqnarray}
\frac{d}{dt}\rho=-\frac{i}{\hbar}[H,\rho]+L_{A}(\rho)+L_{D}(\rho)+L_{\mathrm{DP}}(\rho),
\label{eqs:master_equation}
\end{eqnarray}
where $\rho$ is the density matrix of the molecules and $H$ is the Hamiltonian of the system, given by
\begin{eqnarray}
H=\sum^{n}_{j=1}\hbar\Omega_{j}\sigma_{j}^{+}\sigma_{j}^{-}+\sum^{n}_{j<k}\hbar J_{jk}(\sigma_{j}^{+}\sigma_{k}^{-}+\sigma_{k}^{+}\sigma_{j}^{-}).
\end{eqnarray}
Here, $\sigma_{j}^{+}$ and $\sigma_{j}^{-}$ are the raising and lowering operator for molecule $j$, $\hbar\Omega_{j}$ is the excited energy of the molecule and $J_{jk}$ is the electronic coupling constant between molecules $j$ and $k$. The first and second non-unitary terms $L_{A}(\rho)$ and $L_{D}(\rho)$ describe the processes of absorbing and emitting thermal light and phonons with coupling constants $\eta_{j}$ and $\Gamma_{j}$ at molecule $j$. The second term $L_{D}(\rho)$ also contains an irreversible decay from the receiver (denoted by $j=1$) to the RC with a coupling constant $\Gamma_{\mathrm{RC}}$. The Lindblad operators are given by
\begin{eqnarray}
&L_{A}(\rho)=\sum^{n}_{j=1}\alpha_{j}(\sigma_{j}^{+}\rho\sigma_{j}^{-}-\frac{1}{2}\{\sigma_{j}^{-}\sigma_{j}^{+},\rho\}),\\
&L_{D}(\rho)=\sum^{n}_{j=1}\beta_{j}(\sigma_{j}^{-}\rho\sigma_{j}^{+}-\frac{1}{2}\{\sigma_{j}^{+}\sigma_{j}^{-},\rho\}),
\end{eqnarray}
where $\alpha_{j}=\bar{N}_{l}\eta_{j}+\bar{N}_{p}\Gamma_{j}$ and $\beta_{j}=(\bar{N}_{l}+1)\eta_{j}+(\bar{N}_{p}+1)\Gamma_{j}+\Gamma_{\mathrm{RC}}\delta_{1j}$ are the exciton-creation and exciton-decay constants of molecule $j$. Here, $\bar{N}_{l}$ ($\bar{N}_{p}$) is the mean photon (phonon) number of the thermal light (phonons) and $\delta_{ij}$ is the Kronecker delta with $\delta_{1j}$ indicating that the RC is coupled with molecule 1. The last non-unitary term $L_{\mathrm{DP}}(\rho)$ describes the dephasing process, given by
\begin{eqnarray}
L_{\mathrm{DP}}(\rho)=\sum^{n}_{j=1}\gamma_{j}(\sigma_{j}^{+}\sigma_{j}^{-}\rho\sigma_{j}^{+}\sigma_{j}^{-}-\frac{1}{2}\{\sigma_{j}^{+}\sigma_{j}^{-},\rho\}),
\end{eqnarray}
where $\gamma_{j}$ is the dephasing constant of molecule $j$. This process is also caused by the interaction with phonons with conserving the system energy. By these non-unitary processes, the system state decoheres. To investigate continuous exciton transfer, we consider the steady state of the master equation~\eref{eqs:master_equation}, which is invariant as time evolves. To clarify the principal mechanism, we begin with a single-pathway complex consisting of two antenna molecules and then consider a bi-pathway complex of three antenna molecules (see \fref{fig:antenna_complexes}).

\section{Single-Pathway Complex}

\subsection{Steady Exciton Transfer}

We shall express the steady state condition $\frac{d}{dt}\rho=0$ in terms of the probabilities by eliminating the coherence terms. For the single-pathway complex in \fref{fig:antenna_complexes}(a), the steady state is given in the form of
\begin{eqnarray}
\fl \rho=P_{0}\ket{0}\bra{0}+P_{1}\ket{1}\bra{1}+P_{2}\ket{2}\bra{2}+C_{12}\ket{1}\bra{2}+C_{12}^{*}\ket{2}\bra{1}+P_{12}\ket{12}\bra{12},
\end{eqnarray}
in localized exciton basis $\{\ket{0},\ket{j}\equiv\sigma_{j}^{+}\ket{0},\ket{12}\equiv\sigma_{1}^{+}\sigma_{2}^{+}\ket{0}\}$, where $\ket{0}$ is the ground state of the system. Through the master equation~\eref{eqs:master_equation}, the steady state condition $\frac{d}{dt}\rho=0$ results in the set of equations
\begin{eqnarray}
&\frac{dP_{0}}{dt}=-(\alpha_{1}+\alpha_{2})P_{0}+\beta_{1}P_{1}+\beta_{2}P_{2}=0,\\
&\frac{dP_{1}}{dt}=\alpha_{1}P_{0}-(\alpha_{2}+\beta_{1})P_{1}+i J_{12}(C_{12}-C_{12}^{*})+\beta_{2}P_{12}=0,\label{eq:P1}\\
&\frac{dP_{2}}{dt}=\alpha_{2}P_{0}-(\alpha_{1}+\beta_{2})P_{2}-i J_{12}(C_{12}-C_{12}^{*})+\beta_{1}P_{12}=0,\label{eq:P2}\\
&\frac{dC_{12}}{dt}=iJ_{12}(P_{1}-P_{2})-(D_{12}+i\Delta\Omega_{12})C_{12}=0,\label{eq:coherence1}\\
&\frac{dC_{12}^{*}}{dt}=-iJ_{12}(P_{1}-P_{2})-(D_{12}-i\Delta\Omega_{12})C_{12}^{*}=0,\label{eq:coherence2}\\
&\frac{dP_{12}}{dt}=\alpha_{2}P_{1}+\alpha_{1}P_{2}-(\beta_{1}+\beta_{2})P_{12}=0,
\end{eqnarray}
where $D_{12}=\frac{1}{2}(\alpha_{1}+\beta_{1}+\gamma_{1}+\alpha_{2}+\beta_{2}+\gamma_{2})$ and $\Delta\Omega_{12}=\Omega_{1}-\Omega_{2}$.  The equations for the coherence terms, equations~\eref{eq:coherence1} and \eref{eq:coherence2}, lead to
\begin{eqnarray}
&C_{12}=\frac{iJ_{12}}{D_{12}+i\Delta\Omega_{12}}(P_{1}-P_{2}),\label{eq:sub1}\\
&C_{12}^{*}=\frac{-iJ_{12}}{D_{12}-i\Delta\Omega_{12}}(P_{1}-P_{2})\label{eq:sub2}.
\end{eqnarray}
By substituting equations~\eref{eq:sub1} and \eref{eq:sub2} into equations~\eref{eq:P1} and \eref{eq:P2}, the steady state condition $\frac{d}{dt}\rho=0$ can be expressed in terms of probabilities
\begin{eqnarray}
&\frac{dP_{0}}{dt}=-(\alpha_{1}+\alpha_{2})P_{0}+\beta_{1}P_{1}+\beta_{2}P_{2}=0,\label{eq:st1}\\
&\frac{dP_{1}}{dt}=\alpha_{1}P_{0}-(\alpha_{2}+\beta_{1})P_{1}+\xi_{12}(P_{2}-P_{1})+\beta_{2}P_{12}=0,\label{eq:st2}\\
&\frac{dP_{2}}{dt}=\alpha_{2}P_{0}-(\alpha_{1}+\beta_{2})P_{2}+\xi_{12}(P_{1}-P_{2})+\beta_{1}P_{12}=0,\label{eq:st3}\\
&\frac{dP_{12}}{dt}=\alpha_{2}P_{1}+\alpha_{1}P_{2}-(\beta_{1}+\beta_{2})P_{12}=0,\label{eq:st4}
\end{eqnarray}
where the hopping constant $\xi_{12}$ is defined by $\xi_{12}=\displaystyle\frac{2J_{12}^{2}D_{12}}{D_{12}^{2}+\Delta\Omega_{12}^{2}}$. Focused on $\frac{d}{dt}P_{j}$, $\alpha_{j}P_{0}$ ($\beta_{j}P_{j}$) is the exciton-increasing (decreasing) rate of $P_{j}$ from (to) the ground state, and $\beta_{k}P_{12}$ ($\alpha_{k}P_{j}$) is the converting rate of $P_{j}$ from (to) the two-exciton state. In addition, $\xi_{12}P_{k}$ is the exciton hopping rate from molecule $k$ to molecule $j$, while $\xi_{12}P_{j}$ is that from molecule $j$ to molecule $k$. It is notable that the hopping constant $\xi_{12}$ contains significant information, such as electronic coupling constant $J_{12}$, energy-level mismatch between the donor and receiver $\Delta\Omega_{12}$ and dephasing constants $\gamma_{j}$. The solution to equations \eref{eq:st1}--\eref{eq:st4} is given, with the normalization $\mathrm{Tr}(\rho)=P_{0}+P_{1}+P_{2}+P_{12}=1$, as
\begin{eqnarray}
\fl &P_{0}=\frac{\beta_{1}\beta_{2}(\alpha_{1}+\alpha_{2}+\beta_{1}+\beta_{2})+(\beta_{1}+\beta_{2})^{2}\xi_{12}}{(\alpha_{1}+\alpha_{2}+\beta_{1}+\beta_{2})[(\alpha_{1}+\beta_{1})(\alpha_{2}+\beta_{2})+(\alpha_{1}+\alpha_{2}+\beta_{1}+\beta_{2})\xi_{12}]},\label{solP0}\\
\fl &P_{1}=\frac{\alpha_{1}\beta_{2}(\alpha_{1}+\alpha_{2}+\beta_{1}+\beta_{2})+(\alpha_{1}+\alpha_{2})(\beta_{1}+\beta_{2})\xi_{12}}{(\alpha_{1}+\alpha_{2}+\beta_{1}+\beta_{2})[(\alpha_{1}+\beta_{1})(\alpha_{2}+\beta_{2})+(\alpha_{1}+\alpha_{2}+\beta_{1}+\beta_{2})\xi_{12}]},\label{solP1}\\
\fl &P_{2}=\frac{\alpha_{2}\beta_{1}(\alpha_{1}+\alpha_{2}+\beta_{1}+\beta_{2})+(\alpha_{1}+\alpha_{2})(\beta_{1}+\beta_{2})\xi_{12}}{(\alpha_{1}+\alpha_{2}+\beta_{1}+\beta_{2})[(\alpha_{1}+\beta_{1})(\alpha_{2}+\beta_{2})+(\alpha_{1}+\alpha_{2}+\beta_{1}+\beta_{2})\xi_{12}]},\label{solP2}\\
\fl &P_{12}=\frac{\alpha_{1}\alpha_{2}(\alpha_{1}+\alpha_{2}+\beta_{1}+\beta_{2})+(\alpha_{1}+\alpha_{2})^{2}\xi_{12}}{(\alpha_{1}+\alpha_{2}+\beta_{1}+\beta_{2})[(\alpha_{1}+\beta_{1})(\alpha_{2}+\beta_{2})+(\alpha_{1}+\alpha_{2}+\beta_{1}+\beta_{2})\xi_{12}]}.\label{solP12}
\end{eqnarray}

\subsection{Necessary and sufficient condition for the faithful donor}

We shall find the condition in which the antenna molecules faithfully play their roles. An exciton is created at molecule $j$, if unoccupied, by the rate of $R_{A}^{(j)}=\alpha_{j}(1-W_{j})$ where $W_{j}=P_{j}+P_{12}$ is the probability of finding an exciton at molecule $j$, given by
\begin{eqnarray}
&W_{1}=\frac{\alpha_{1}(\alpha_{2}+\beta_{2})+(\alpha_{1}+\alpha_{2})\xi_{12}}{(\alpha_{1}+\beta_{1})(\alpha_{2}+\beta_{2})+(\alpha_{1}+\alpha_{2}+\beta_{1}+\beta_{2})\xi_{12}},\\
&W_{2}=\frac{\alpha_{2}(\alpha_{1}+\beta_{1})+(\alpha_{1}+\alpha_{2})\xi_{12}}{(\alpha_{1}+\beta_{1})(\alpha_{2}+\beta_{2})+(\alpha_{1}+\alpha_{2}+\beta_{1}+\beta_{2})\xi_{12}}.
\end{eqnarray}
On the other hand, molecule $j$ loses its exciton, if occupied, by the rate of $R_{D}^{(j)}=\beta_{j}W_{j}$. Here, $R_{D}^{(1)}$ includes the transfer to the RC that happens by the rate of $R_{\mathrm{RC}}=\Gamma_{\mathrm{RC}}W_{1}$. The net rate of exciton hopping from the donor to the receiver $R_{H}^{(12)}\equiv\xi_{12}(P_{2}-P_{1})$ is given by 
\begin{eqnarray}
R_{H}^{(12)}=\frac{(\alpha_{2}\beta_{1}-\alpha_{1}\beta_{2})\xi_{12}}{(\alpha_{1}+\beta_{1})(\alpha_{2}+\beta_{2})+(\alpha_{1}+\alpha_{2}+\beta_{1}+\beta_{2})\xi_{12}},
\end{eqnarray}
so that we have two rate equations: $R_{A}^{(1)}+R_{H}^{(12)}=R_{D}^{(1)}$ and $R_{A}^{(2)}=R_{D}^{(2)}+R_{H}^{(12)}$.

Now we consider the effect of attaching the donor to the receiver ($\xi_{12}>0$ or equivalently $J_{12}\neq 0$) compared to the decoupled system ($J_{12}=0$). The attachment increases the net absorption rate of light (and also phonons) $R_{A}=R_{A}^{(1)}+R_{A}^{(2)}$, as seen from the positivity of the derivative of the net absorption rate $R_{A}$ with respect to the hopping constant $\xi_{12}$
\begin{eqnarray}
	\frac{\partial R_{A}}{\partial \xi_{12}}=\left[\frac{\alpha_{2}\beta_{1}-\alpha_{1}\beta_{2}}{(\alpha_{1}+\beta_{1})(\alpha_{2}+\beta_{2})+(\alpha_{1}+\alpha_{2}+\beta_{1}+\beta_{2})\xi_{12}}\right]^{2}>0.
\end{eqnarray}
One might conjecture that the attachment also increases the exciton transfer rate to the RC, but this is not necessarily the case. In fact, the transfer rate is enhanced if and only if
\begin{eqnarray}
\frac{\partial R_{\mathrm{RC}}}{\partial \xi_{12}}=\frac{\Gamma_{\mathrm{RC}}(\alpha_{2}+\beta_{2})(\alpha_{2}\beta_{1}-\alpha_{1}\beta_{2})}{[(\alpha_{1}+\beta_{1})(\alpha_{2}+\beta_{2})+(\alpha_{1}+\alpha_{2}+\beta_{1}+\beta_{2})\xi_{12}]^{2}}>0,
\end{eqnarray}
or equivalently
\begin{eqnarray}
\mathcal{A}_{2}>\mathcal{A}_{1},
\label{eqs:enhancement_condition}
\end{eqnarray}
where $\mathcal{A}_{j}=\alpha_{j}/\beta_{j}$ is the effective absorption ratio of molecule $j$. This condition holds if $\Gamma_{\mathrm{RC}}$ is large enough as $\mathcal{A}_{1}$ is proportional to $\Gamma_{\mathrm{RC}}^{-1}$. This is also the condition that the net exciton hopping is directed from the donor to the receiver ($R_{H}^{(12)}>0$). Thus, {\it attaching a donor in the condition~\eref{eqs:enhancement_condition} faithfully directs the energy flow from the antenna molecules eventually to the RC as well as increasing the net absorption rate of light}. If the condition~\eref{eqs:enhancement_condition} is not satisfied, the net exciton hopping is directed from the receiver to the donor ($R_{H}^{(12)}<0$) so that removing a donor from the antenna complex ($J_{12}=0$) increases the exciton transfer rate to the RC. These results are clearly reflected in the transfer efficiency $\epsilon\equiv R_{\mathrm{RC}}/R_{A}\in[0,1]$, given as
\begin{eqnarray}
	\epsilon=\frac{\Gamma_{\mathrm{RC}}}{\beta_{\mathrm{total}}}\left[1+\displaystyle\frac{\beta_{2}}{\beta_{\mathrm{total}}}\frac{\alpha_{2}\beta_{1}-\alpha_{1}\beta_{2}}{\xi_{12}(\alpha_{1}+\alpha_{2})+\alpha_{1}(\alpha_{2}+\beta_{2})}\right]^{-1},
\end{eqnarray}
where $\beta_{\mathrm{total}}=\sum^{n}_{j=1}\beta_{j}$ is the total exciton-decay constant of the antenna complex. We would note that the attachment decreases the receiver's absorption rate $R_{A}^{(1)}$ while increasing the donor's $R_{A}^{(2)}$, thus reducing the role of receiver in the absorption and changing it to the transmission to the RC. Under physiological conditions, the intensity of sunlight is weak ($\alpha_{j}\ll\beta_{k}$) and the single-exciton manifold is of primary importance for modeling photosynthetic complexes such as the  Fenna-Matthews-Olson (FMO) complex~\cite{Cheng2009,Ishizaki2010}. In this case, the steady state is well-approximated within the single-exciton manifold and $\epsilon$ is reduced to the single-exciton transfer efficiency~\cite{Mohseni2008}--\cite{Rebentrost2009},
\begin{eqnarray}
\epsilon=\Gamma_{\mathrm{RC}}\int^{\infty}_{0}dt\bra{1}\rho(t)\ket{1},
\end{eqnarray}
where the initial state is given by $\rho(0)=\sum^{n}_{j=1}(\alpha_{j}/\alpha_{\mathrm{total}})\ket{j}\bra{j}$ with $\alpha_{\mathrm{total}}=\sum^{n}_{j=1}\alpha_{j}$ and its dynamics is governed by the master equation in the form of
\begin{eqnarray}
\frac{d}{dt}\rho=-\frac{i}{\hbar}[H,\rho]+L_{D}(\rho)+L_{\mathrm{DP}}(\rho).
\label{eq:single_exciton_master_equation}
\end{eqnarray}

\subsection{Subsidiary Role of Dephasing Process}

\begin{figure} [t]
\centering
\subfigure{\includegraphics[width=0.55\textwidth]{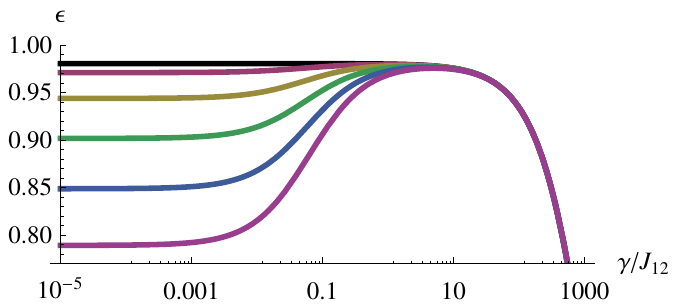}}
\subfigure{\includegraphics[width=0.11\textwidth]{figure2b-eps-converted-to.pdf}}
\caption{(color online) Transfer efficiency $\epsilon$ of \fref{fig:antenna_complexes}(a) under low energy absorption ($\alpha_{1}=0$, $\alpha_{2}\ll\beta_{2}$). We choose $\Gamma_{\mathrm{RC}}/J_{12}=10^{-1}$, $\Gamma_{j}/J_{12}=10^{-3}$ ($\beta_{j}=\Gamma_{j}+\Gamma_{\mathrm{RC}}\delta_{1j}$), $\gamma_{j}/J_{12}=\gamma/J_{12}$ and $\Delta\Omega_{12}/J_{12}=k$ for $k=0,1,\cdots,5$. The region of large $\gamma/J_{12}$ is only for eye-guiding.}
\label{fig:single_pathway_complex}
\end{figure}

We will consider the noise effect on the transfer efficiency $\epsilon$ with respect to the degree of the energy-level mismatch, as shown in \fref{fig:single_pathway_complex}. Assume that the condition of enhancement for the transfer efficiency, Eq.~(\ref{eqs:enhancement_condition}) holds. Adding dephasing noise increases $\epsilon$ when the energy-level mismatch $\abs{\Delta\Omega_{12}}$ is larger than the constant $D_{12}$ of no dephasing $\gamma_{j}=0$. However, $\epsilon$ is still maximized when there are no energy-level mismatch and no dephasing ($\Delta\Omega_{12}=\gamma_{j}=0$). The noise, if present, never makes any enhancement.

These results remain unaltered for a single-pathway complex consisting of more than two molecules. Plenio {\it et al.}~\cite{Plenio2008} investigated the transfer efficiency for a single-pathway complex of a uniform linear chain where all molecules have the same energy level, and their electronic couplings and exciton-decay rates are uniform. Assuming that the initial excitation is located at an edge of the chain and the receiver molecule is located at the opposite one, they numerically observed that the dephasing does not improve the transfer efficiency. The observation implies that there is no Anderson localization effect and the dephasing noise never enhances the transfer efficiency if any energy-level mismatches are absent. Large energy-level mismatches cause the Anderson localization and the dephasing noise releases the inhibition due to the localization, leading to the enhancement of the transfer efficiency. However, the optimal structure of a single-pathway complex is that with no energy-level mismatches. In this sense, the noise-assisted enhancement of $\epsilon$ is a subsidiary effect once the energy-level mismatch settles in the system. In the next section, we show that the negative role of energy-level mismatches and the subsidiary role of dephasing noise dramatically change in a multi-pathway complex.

\section{Bi-Pathway Complex}

\subsection{Enhancement of Transfer Efficiency by Energy-Level Mismatches}

Consider a bi-pathway complex as in \fref{fig:antenna_complexes}(b), where two identical donors ($\alpha_{d}=\alpha_{2}=\alpha_{3}$, $\beta_{d}=\beta_{2}=\beta_{3}$, $\gamma_{d}=\gamma_{2}=\gamma_{3}$) are independently coupled to a receiver molecule ($J_{23}=0$). We assume low energy absorption ($\alpha_{j}\ll\beta_{k}$), taking the single-exciton approximation. We also take $\alpha_{1}=0$, reminding that attaching donors transforms the receiver's role to a transmission channel in the single-pathway complex~\footnote{In both cases of \fref{fig:antenna_complexes}(a) and (b), under the same condition~(\ref{eqs:enhancement_condition}), the net exciton transfer is directed from the donor(s) to the receiver.}. In this case, the transfer efficiency $\epsilon$ is reduced to the single-exciton transfer efficiency when the initial state is given by $\frac{1}{2}(\ket{2}\bra{2}+\ket{3}\bra{3})$:
\begin{eqnarray}
	\epsilon=\frac{\Gamma_{\mathrm{RC}}}{\beta_{\mathrm{total}}}\left[1+\displaystyle\frac{\beta_{1}\beta_{d}}{\beta_{\mathrm{total}}}\frac{\frac{1}{2}(\xi_{12}+\xi_{13})+2(\xi_{23}+\frac{1}{2}\beta_{d})}{\xi_{12}\xi_{13}+(\xi_{12}+\xi_{13})(\xi_{23}+\frac{1}{2}\beta_{d})}\right]^{-1},
\label{eq:bieff}
\end{eqnarray}
where $\beta_{\mathrm{total}}=\beta_{1}+\beta_{2}+\beta_{3}$. Here, the hopping constants $\xi_{jk}$ between molecules $j$ and $k$ are given by
\begin{eqnarray}
	&\xi_{12}=\frac{J_{12}^{2}(J_{12}^{2}-J_{13}^{2}+S_{13}S_{23})}{J_{12}^{2}S_{21}+J_{13}^{2}S_{13}+S_{21}S_{13}S_{23}}+{\it c.c.},\label{eq:xi12}\\
	&\xi_{13}=\frac{J_{13}^{2}(J_{13}^{2}-J_{12}^{2}+S_{12}S_{32})}{J_{13}^{2}S_{31}+J_{12}^{2}S_{12}+S_{31}S_{12}S_{32}}+{\it c.c.},\\
	&\xi_{23}=\frac{J_{12}^{2}J_{13}^{2}}{J_{12}^{2}S_{21}+J_{13}^{2}S_{13}+S_{21}S_{13}S_{23}}+{\it c.c.}\label{eq:xi23},
\end{eqnarray}
where {\it c.c.} stands for the complex conjugate, $S_{jk}=D_{jk}+i\Delta\Omega_{jk}$, $D_{jk}=\frac{1}{2}(\beta_{j}+\gamma_{j}+\beta_{k}+\gamma_{k})$ and $\Delta\Omega_{jk}=\Omega_{j}-\Omega_{k}$. Note that $\epsilon$ is a monotonically increasing function of the hopping constants $\xi_{jk}$: $\displaystyle\frac{\partial \epsilon}{\partial \xi_{jk}}>0$, $\forall j,k$.

\begin{figure} [t]
\centering
\subfigure[No dephasing noise]{\includegraphics{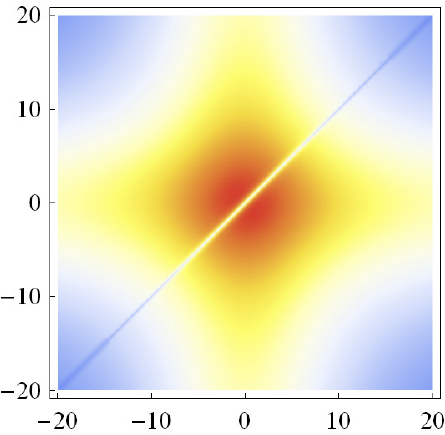}}\hspace{10pt}
\subfigure[Dephasing at the receiver]{\includegraphics{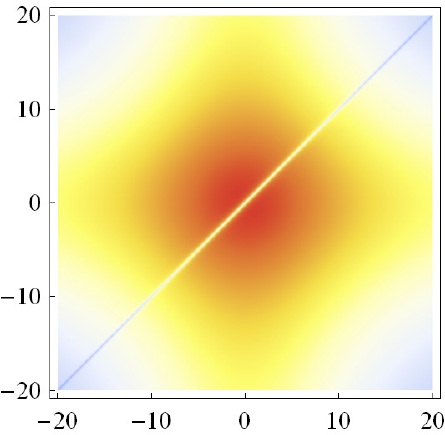}}\hspace{10pt}
\subfigure[Dephasing at the donors]{\includegraphics{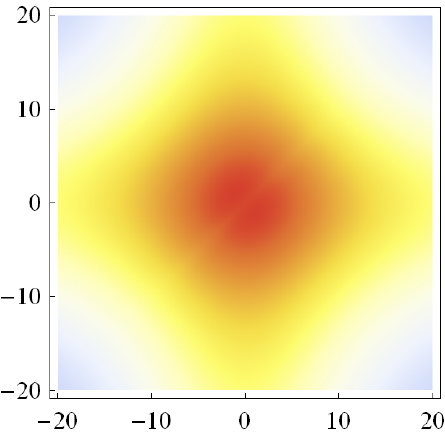}}\\
\subfigure[No dephasing noise]{\includegraphics{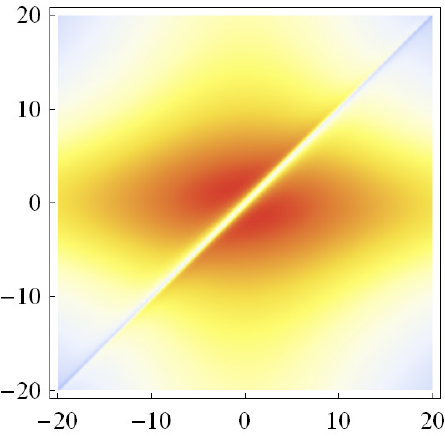}}\hspace{10pt}
\subfigure[Dephasing at the receiver]{\includegraphics{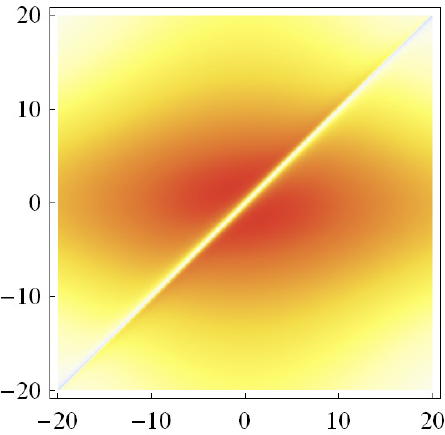}}\hspace{10pt}
\subfigure[Dephasing at the donors]{\includegraphics{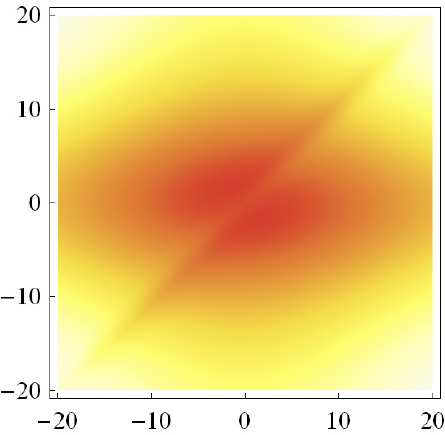}}\\
\subfigure[No dephasing noise]{\includegraphics{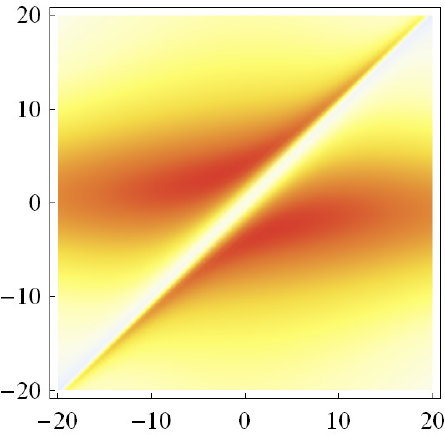}}\hspace{10pt}
\subfigure[Dephasing at the receiver]{\includegraphics{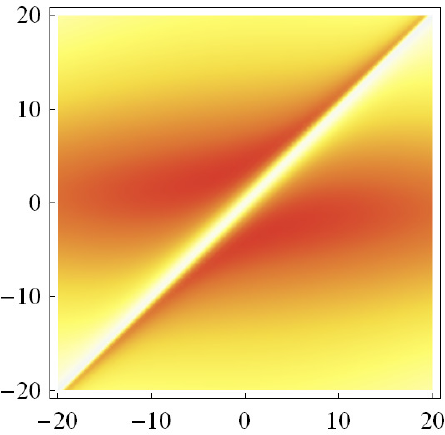}}\hspace{10pt}
\subfigure[Dephasing at the donors]{\includegraphics{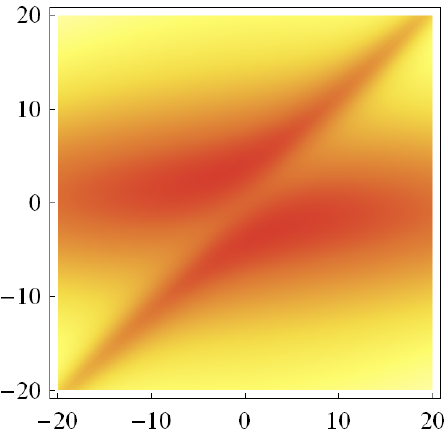}}\\
\vspace{10pt}
\hspace{235pt}
\subfigure{\includegraphics[width=0.4\textwidth]{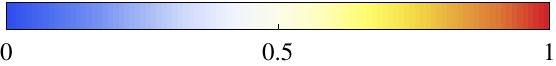}}
\caption{(color online) Transfer efficiency $\epsilon$ of the bi-pathway complex in \fref{fig:antenna_complexes}(b), parameterized by energy-level mismatches. In each panel, horizontal and vertical axes represent energy-level mismatches $\Delta\Omega_{12}/J_{13}$ and $\Delta\Omega_{13}/J_{13}$, respectively. From first to third rows, we take $J_{12}/J_{13}=1,2,4$, respectively. In both cases, the parameters are chosen as $\Gamma_{\mathrm{RC}}/J_{13}=10^{-1}$, $\Gamma_{j}/J_{13}=10^{-3}$ ($\beta_{j}=\Gamma_{j}+\Gamma_{\mathrm{RC}}\delta_{1j}$) and the dephasing constants $(\gamma_{1}/J_{13},\gamma_{d}/J_{13})$ = $(0,0)$, $(10^{-1},0)$, $(0,10^{-1})$ from the left to the right.}
\label{fig:energy_mismatches}
\end{figure}

The transfer efficiency $\epsilon$ is presented as a function of energy-level mismatches in \fref{fig:energy_mismatches}. Note that in each panel, {\it near-unit transfer efficiency is located at a wide range of energy-level mismatches and the maximal transfer efficiency is located on asymmetric energy-level mismatches $\Delta\Omega_{12}\neq\Delta\Omega_{13}$}. In the first column of panels where no dephasing is assumed, the effect of energy-level mismatches are clearly shown: Symmetric energy-level mismatches $\Delta\Omega_{12}=\Delta\Omega_{13}$ result in the suppression of $\epsilon<1/2$, whereas asymmetric ones lead to the near-unit transfer efficiency. As in the second column of panels, adding dephasing $\gamma_{1}$ at the receiver can slightly enhance $\epsilon$, but the previous tendency by the energy-level mismatches is rather unaltered: For the symmetric energy-level mismatches $\Delta\Omega=\Delta\Omega_{12}=\Delta\Omega_{13}$, in the absence of dephasing noise at the donors ($\gamma_{d}=0$), the transfer efficiency is reduced to
\begin{equation}
	\epsilon=\frac{\Gamma_{\mathrm{RC}}}{\beta_{1}+\beta_{d}}\left[2+\frac{\beta_{1}\beta_{d}}{(\beta_{1}+\beta_{d})(J_{12}^{2}+J_{13}^{2})}\frac{D_{1d}^{2}+\Delta\Omega^{2}}{D_{1d}}\right]^{-1}<\frac{1}{2},
\label{eq:seproof}
\end{equation}
where $D_{1d}=\frac{1}{2}(\beta_{1}+\gamma_{1}+\beta_{d})$. However, it changes dramatically when the donors are under dephasing noise. Increasing dephasing $\gamma_{d}$ at the donors results in the great improvement so that $\epsilon$ goes over $1/2$ near to the unity even for the symmetric energy-level mismatches. These results remain valid for asymmetric coupling constants $J_{12}\neq J_{13}$, as seen by comparing three rows in \fref{fig:energy_mismatches}.

\subsection{Destructive Interference at the Receiver Molecule}

\begin{figure} [t]
\subfigure[Molecule 1 (Receiver)]{\includegraphics[width=0.333\textwidth]{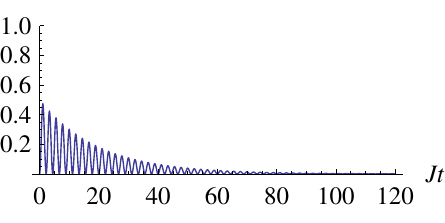}}\hspace{3pt}
\subfigure[Molecule 2 (Donor)]{\includegraphics[width=0.333\textwidth]{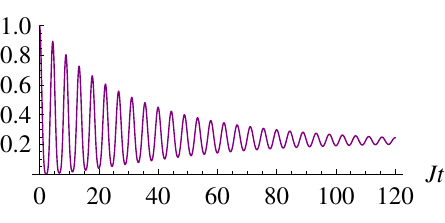}}\hspace{3pt}
\subfigure[Molecule 3]{\includegraphics[width=0.333\textwidth]{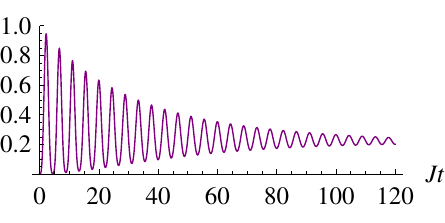}}
\caption{Population relaxations of molecules 1, 2 and 3 with respect to time in the bi-pathway complex when there are no energy-level mismatches and no dephasing noise, but the decay constants $\Gamma_{\mathrm{RC}}/J=10^{-1}$ and $\Gamma_{j}/J=10^{-3}$ ($\beta_{j}=\Gamma_{j}+\Gamma_{\mathrm{RC}}\delta_{1j}$). An exciton created at molecule 2 hops over the receiver and donor molecules, and eventually relaxes in two different ways: Either the exciton is transferred through the molecule 1 (receiver) to the RC, or it relaxes at the donors to the environment. (a) The population at the receiver molecule becomes negligible for $Jt>60$, whereas (b)/(c) the populations at the donors last even beyond $60$ units of $Jt$. These results imply the evidence of the destructive interference at the receiver molecule. Due to the destructive interference, since about $60$ units of $Jt$, the exciton is trapped at the donors and consequently relaxes to the environment.}
\label{fig:population_dynamics}
\end{figure}

{\it The underlying mechanism is the quantum destructive interference at the receiver molecule and its suppression by the energy-level mismatches and/or the dephasing noise}. In the presence of energy-decay process, once excited by light absorption, a localized state $\ket{j}$ evolves into a statistical mixture, $(1-p_{j}(t))\ket{0}\bra{0}+p_{j}(t)\rho_{j}(t)$, of the ground state and a delocalized state $\rho_{j}(t)$. Here, the delocalized state is a superposition of localized single-exciton states. In the absence of dephasing noise, the delocalized state $\rho_{j}(t)$ remains in a pure state $p_{j}(t)\rho_{j}(t)=\ket{\psi_{j}(t)}\bra{\psi_{j}(t)}$ for which it is convenient to employ a stochastic Schr\"odinger equation~\cite{Gardiner1991,Breuer2002},
\begin{eqnarray}
i\hbar\frac{d}{dt}\ket{\psi_{j}(t)}=K\ket{\psi_{j}(t)}\quad \mathrm{with}\quad K=H-i\hbar\sum^{n}_{k=1}\frac{1}{2}\beta_{k}\sigma_{k}^{+}\sigma_{k}^{-}.
\label{eq:stochasticK}
\end{eqnarray}
Here, $\ket{\psi_{j}(t)}$ is unnormalized and $p_{j}(t)=\ave{\psi_{j}(t)|\psi_{j}(t)}$ is the probability that an exciton is still present in the molecules. For the symmetric energy-level mismatches $\Delta\Omega=\Delta\Omega_{12}=\Delta\Omega_{13}$, the delocalized state is written as
\begin{eqnarray}
\ket{\psi_{2}(t)}=\frac{J_{13}}{M_{1}}\exp(\frac{E_{1}}{i\hbar}t)\ket{v_{1}}+\sum^{3}_{p=2}\frac{J_{12}}{M_{p}}\exp(\frac{E_{p}}{i\hbar}t)\ket{v_{p}},\label{eq:pure_destructive_interference}
\end{eqnarray}
\begin{eqnarray}
\ket{\psi_{3}(t)}=-\frac{J_{12}}{M_{1}}\exp(\frac{E_{1}}{i\hbar}t)\ket{v_{1}}+\sum^{3}_{p=2}\frac{J_{13}}{M_{p}}\exp(\frac{E_{p}}{i\hbar}t)\ket{v_{p}},\label{eq:pure_destructive_interference_2}
\end{eqnarray}
where $\ket{\psi_{j}(0)}=\ket{j}$ for $j=2,3$. Here, $\ket{v_{p}}$ are the right eigenvectors of $K$, given as
\begin{eqnarray}
&\ket{v_{1}}=J_{13}\ket{2}-J_{12}\ket{3},\label{eq:v1}\\
&\ket{v_{p}}=-\frac{\Delta E_{1p}}{\hbar}\ket{1}+J_{12}\ket{2}+J_{13}\ket{3},\quad p\neq 1,
\end{eqnarray}
where $\Delta E_{1p}=E_{1}-E_{p}$, $M_{p}=J_{12}^{2}+J_{13}^{2}+(\Delta E_{1p}/\hbar)^{2}$ and $E_{p}$ are the complex eigenvalues corresponding to $\ket{v_{p}}$, given by
\begin{eqnarray}
&E_{1}=i\hbar(-i(\Omega_{1}-\Delta\Omega)-\frac{1}{2}\beta_{d}),\label{eq:evalue1}\\
&E_{p}=i\hbar(X-(-1)^{p}Y),\quad p\neq 1,\label{eq:evalue23}
\end{eqnarray}
where $X=-i(\Omega_{1}-\frac{1}{2}\Delta\Omega)-\frac{1}{4}(\beta_{1}+\beta_{d})$, $Y=\frac{1}{4}\sqrt{(\beta_{1}-\beta_{d}+2i\Delta\Omega)^{2}-16(J_{12}^{2}+J_{13}^{2})}$. Note that when normalized, $\ket{v_{1}}$ has no probability amplitude of staying in the receiver due to the perfect destructive interference between the amplitudes of multiple paths to and from the donors. Thus, $\ket{v_{1}}$ does not contribute to the energy transfer to the RC, so that it states the eventual loss of excitons at the donors. The destructive interference consequently results in the transfer efficiency less than $1/2$ (see \fref{fig:population_dynamics}). For the asymmetric energy-level mismatches, on the other hand, the destructive interference diminishes as every eigenvector of $K$ has amplitude at the receiver. The transfer efficiency increases as biasing the asymmetric degree of energy-level mismatches and turns to decrease for large mismatches (see \fref{fig:energy_mismatches}). The destructive interference can also be avoided by the dephasing at the donors that collapses $\ket{v_{1}}$ probabilistically into one of the localized states at the donors, which have a chance to transfer an exciton to the RC.

\subsection{Quantitative Description of Noise-Assisted Enhancement}

Quantitative description of the noise-assisted enhancement can also be provided by using a stochastic Schr\"odinger equation. Assume a localized state $\ket{j}$ is created at time $t=0$. In the presence of the exciton-decay and dephasing processes, the localized state $\ket{j}$ evolves into a statistical mixture of the ground state and a delocalized state,
\begin{eqnarray}
\rho=(1-p_{j}(t))\ket{0}\bra{0}+p_{j}(t)\rho_{j}(t).
\end{eqnarray}
Here, $(1-p_{j}(t))$ and $p_{j}(t)$ are the probabilities for finding the system in the ground state and the delocalized state, respectively. The delocalized state $\rho_{j}(t)$ is a statistical mixture of single-exciton states. As time evolves, the delocalized state $\rho_{j}(t)$ is probabilistically collapsed into the ground state $\ket{0}$ by the exciton-decay process at molecule $k$ with the rate of $\beta_{k}\bra{k}\rho(t)\ket{k}$. The delocalized state $\rho_{j}(t)$ is also probabilistically collapsed into the localized state $\ket{k}$ by the dephasing process at molecule $k$ with the rate of $\gamma_{k}\bra{k}\rho(t)\ket{k}$. These stochastic processes are equivalent to the master equation~\eref{eq:single_exciton_master_equation} that can be rearranged as
\begin{eqnarray}
\frac{d}{dt}\rho=-\frac{i}{\hbar}(K\rho-\rho K^{\dagger})+\sum^{n}_{k=1}\beta_{k}\sigma_{k}^{-}\rho\sigma_{k}^{+}+\sum^{n}_{k=1}\gamma_{k}\sigma_{k}^{+}\sigma_{k}^{-}\rho\sigma_{k}^{+}\sigma_{k}^{-},
\end{eqnarray}
where $K=H-i\hbar\sum^{n}_{k=1}\frac{1}{2}(\beta_{k}+\gamma_{k})\sigma_{k}^{+}\sigma_{k}^{-}$. To describe the dynamics of the state $\rho$ based on the stochastic processes, we express the delocalized state $\rho_{j}(t)$ as a statistical mixture of pure single-exciton states
\begin{eqnarray}
\fl p_{j}(t)\rho_{j}(t)=&\ket{\psi_{j}(t;0)}\bra{\psi_{j}(t;0)}+\sum_{k=1}^{n}\int^{t}_{0}dt_{1}\ket{\psi_{k}(t;t_{1})}\bra{\psi_{k}(t;t_{1})}R_{kj}^{\mathrm{(DP)}}(t_{1};0)\nonumber\\
\fl &+\sum_{k,l=1}^{n}\int^{t}_{t_{1}}dt_{2}\int^{t}_{0}dt_{1}\ket{\psi_{l}(t;t_{2})}\bra{\psi_{l}(t;t_{2})}R_{lk}^{\mathrm{(DP)}}(t_{2};t_{1})R_{kj}^{\mathrm{(DP)}}(t_{1};0)+\cdots.
\end{eqnarray}
Here, $\ket{\psi_{k}(t;t_{z})}$ is an unnormalized pure state such that $\ket{\psi_{k}(t;t_{z})}=0$ for time $t<t_{z}$, $\ket{\psi_{k}(t;t_{z})}=\ket{k}$ at $t=t_{z}$ and the dynamics of $\ket{\psi_{k}(t;t_{z})}$ for $t\geq t_{z}$ is determined by a stochastic Schr\"odinger equation~\cite{Gardiner1991,Breuer2002} in the form of
\begin{eqnarray}
i\hbar\frac{d}{dt}\ket{\psi_{k}(t;t_{z})}=K\ket{\psi_{k}(t;t_{z})}.
\end{eqnarray}
The formal solution is given as $\ket{\psi_{k}(t;t_{z})}=\exp[-\frac{i}{\hbar}K(t-t_{z})]\ket{k}=\ket{\psi_{k}(t-t_{z};0)}$ for $t\geq t_{z}$. $R_{lk}^{\mathrm{(DP)}}(t_{z+1};t_{z})=\gamma_{l}\abs{\ave{l|\psi_{k}(t_{z+1};t_{z})}}^{2}$ is the rate that $\ket{\psi_{k}(t;t_{z})}$ is collapsed into the localized state $\ket{l}$ by the dephasing process at $t=t_{z+1}$. The probability to be in the ground state $\ket{0}$ at time $t$ is given by
\begin{eqnarray}
\fl 1-p_{j}(t)=&\sum_{k=1}^{n}\int^{t}_{0}dt_{1}R_{kj}^{(D)}(t_{1};0)+\sum_{k,l=1}^{n}\int^{t}_{t_{1}}dt_{2}\int^{t}_{0}dt_{1}R_{lk}^{(D)}(t_{2};t_{1})R_{kj}^{\mathrm{(DP)}}(t_{1};0)\nonumber\\
\fl &+\sum_{k,l,m=1}^{n}\int^{t}_{t_{2}}dt_{3}\int^{t}_{t_{1}}dt_{2}\int^{t}_{0}dt_{1}R_{ml}^{(D)}(t_{3};t_{2})R_{lk}^{\mathrm{(DP)}}(t_{2};t_{1})R_{kj}^{\mathrm{(DP)}}(t_{1};0)+\cdots,
\end{eqnarray}
where $R_{lk}^{(D)}(t_{z+1};t_{z})=\beta_{l}\abs{\ave{l|\psi_{k}(t_{z+1};t_{z})}}^{2}$ is the rate that $\ket{\psi_{k}(t;t_{z})}$ is collapsed into the ground state by the exciton-decay process at $t=t_{z+1}$.

We shall express the single-exciton transfer efficiency $\epsilon$ in equation~\eref{eq:bieff} as a series sum of the transfer probabilities to the RC on the basis of the stochastic processes. There happen two types of stochastic processes in our model, one is the exciton-decay process and the other is the dephasing process. In order to analyze what processes and how many times the processes are involved in the transfer to the RC, it is convenient to consider the temporally-accumulated stochastic probabilities. A (temporally-accumulated) decay probability is defined by
\begin{eqnarray}
P_{kj}^{(D)}=\int^{\infty}_{t_{z}}dt\beta_{k}\abs{\ave{k|\psi_{j}(t;t_{z})}}^{2}=\int^{\infty}_{0}dt\beta_{k}\abs{\ave{k|\psi_{j}(t;0)}}^{2}\quad \forall z,
\end{eqnarray}
where we used $\ket{\psi_{k}(t;t_{z})}=\ket{\psi_{k}(t-t_{z};0)}$ for $t\geq t_{z}$. This is the transition probability from the state $\ket{j}$ to the ground state $\ket{0}$ conditioned by a decay process at molecule $k$. Similarly, define a (temporally-accumulated) dephasing probability as
\begin{eqnarray}
P_{kj}^{\mathrm{(DP)}}=\int^{\infty}_{t_{z}}dt\gamma_{k}\abs{\ave{k|\psi_{j}(t;t_{z})}}^{2}=\int^{\infty}_{0}dt\gamma_{k}\abs{\ave{k|\psi_{j}(t;0)}}^{2}\quad \forall z,
\end{eqnarray}
which is the transition probability from the state $\ket{j}$ to the state $\ket{k}$ conditioned by a dephasing process at molecule $k$. Here, $P_{1j}^{(D)}$ includes both probabilities of the loss to the environment and of the transfer to the RC through the receiver molecule. The transfer probability to the RC is segregated by $(\Gamma_{\mathrm{RC}}/\beta_{1})P_{1j}^{(D)}$. As there are no other stochastic processes assumed, the probabilities $P_{kj}^{(D)}$ and $P_{kj}^{\mathrm{(DP)}}$ satisfy the normalization condition,
\begin{eqnarray}
\sum_{k=1}^{3}P_{kj}^{(S)}=1,
\end{eqnarray}
where $P_{kj}^{(S)}=P_{kj}^{(D)}+P_{kj}^{\mathrm{(DP)}}=\int^{\infty}_{0}dtD_{kk}\abs{\ave{k|\psi_{j}(t;0)}}^{2}$ and $D_{kk}=\beta_{k}+\gamma_{k}$. Now, the single-exciton transfer efficiency $\epsilon_{j}$ of the initial state $\ket{j}$ is expanded into the sum of the stochastic chains of the decay and dephasing processes:
\begin{eqnarray}
\epsilon_{j}
&=(\Gamma_{\mathrm{RC}}/\beta_{1})\left(P_{1j}^{(D)}+\sum_{k=1}^{3}P_{1k}^{(D)}P_{kj}^{\mathrm{(DP)}}+\sum_{k,l=1}^{3}P_{1l}^{(D)}P_{lk}^{\mathrm{(DP)}}P_{kj}^{\mathrm{(DP)}}+\cdots\right)\label{eq:efficiency_expansion_1}\\
&=(\Gamma_{\mathrm{RC}}/\beta_{1})\left(\sum^{\infty}_{z=0}\boldsymbol{P_{D}}(\boldsymbol{P_{\mathrm{DP}}})^{z}\right)_{1j}.\label{eq:efficiency_expansion_2}
\end{eqnarray}
Here, a single chain $(\Gamma_{\mathrm{RC}}/\beta_{1})(\boldsymbol{P_{D}}(\boldsymbol{P_{\mathrm{DP}}})^{z})_{1j}$ is the transfer probability to the RC of the initial state $\ket{j}$ through $z$ times dephasing processes followed by the decay process. The matrices $\boldsymbol{P_{D}}$ and $\boldsymbol{P_{\mathrm{DP}}}$ of the decay and dephasing probabilities are given, respectively, as
\begin{eqnarray}
\fl \boldsymbol{P_{D}}&=-
\begin{bmatrix}
	\beta_{1} & 0 & 0 \\ 
	0 & \beta_{2} & 0 \\
	0 & 0 & \beta_{3}
\end{bmatrix}
\begin{bmatrix}
	-D_{11}-\xi_{12}-\xi_{13} & \xi_{12} & \xi_{13} \\ 
	\xi_{12} & -D_{22}-\xi_{12}-\xi_{23} & \xi_{23} \\
	\xi_{13} & \xi_{23} & -D_{33}-\xi_{13}-\xi_{23}
\end{bmatrix}^{-1},\\
\fl \boldsymbol{P_{\mathrm{DP}}}&=-
\begin{bmatrix}
	\gamma_{1} & 0 & 0 \\ 
	0 & \gamma_{2} & 0 \\
	0 & 0 & \gamma_{3}
\end{bmatrix}
\begin{bmatrix}
	-D_{11}-\xi_{12}-\xi_{13} & \xi_{12} & \xi_{13} \\ 
	\xi_{12} & -D_{22}-\xi_{12}-\xi_{23} & \xi_{23} \\
	\xi_{13} & \xi_{23} & -D_{33}-\xi_{13}-\xi_{23}
\end{bmatrix}^{-1},
\end{eqnarray}
where $\xi_{jk}$ are the hopping constants defined in equations~\eref{eq:xi12}--\eref{eq:xi23}. The single-exciton transfer efficiency $\epsilon$ in equation~\eref{eq:bieff} is expressed in the form of
\begin{eqnarray}
\epsilon=\sum^{3}_{j=2}\epsilon_{j}\times\frac{1}{2}.
\end{eqnarray}
where $1/2$ is the probability that a single exciton is initially created at molecule $j$.

The stochastic chains can be applied to explain the noise-assisted enhancement. Particularly, consider the case of symmetric energy-level mismatches $\Delta\Omega=\Delta\Omega_{12}=\Delta\Omega_{13}$. The sum of the probabilities that the initial exciton decoheres to the donors (molecule $k=2,3$) once created at the donors (molecule $j=2,3$) in the probabilities $1/2$ is given as
\begin{eqnarray}
\fl P_{dd}^{(S)}\equiv\sum_{j,k=2}^{3}P_{kj}^{(S)}\times\frac{1}{2}=1-\left[2+D_{dd}\left(\frac{2}{D_{11}}+\frac{1}{J_{12}^{2}+J_{13}^{2}}\frac{D_{1d}^{2}+\Delta\Omega^{2}}{D_{1d}}\right)\right]^{-1}>\frac{1}{2},
\end{eqnarray}
where $D_{1d}=\frac{1}{2}(\beta_{1}+\gamma_{1}+\beta_{d}+\gamma_{d})$ and $D_{dd}=\beta_{d}+\gamma_{d}$. Noting $P_{kj}^{(D)}=(\beta_{d}/D_{dd})P_{kj}^{(S)}$ and $P_{kj}^{\mathrm{(DP)}}=(\gamma_{d}/D_{dd})P_{kj}^{(S)}$ for $k=2,3$, the decay and dephasing probabilities at the donors are respectively given as
\begin{eqnarray}
\fl P_{dd}^{(D)}\equiv\sum_{j,k=2}^{3}P_{kj}^{(D)}\times\frac{1}{2}=\frac{\beta_{d}}{D_{dd}}P_{dd}^{(S)},\qquad\qquad
P_{dd}^{\mathrm{(DP)}}\equiv\sum_{j,k=2}^{3}P_{kj}^{\mathrm{(DP)}}\times\frac{1}{2}=\frac{\gamma_{d}}{D_{dd}}P_{dd}^{(S)},
\end{eqnarray}
while those at the receiver molecule are respectively given as
\begin{eqnarray}
\fl P_{1d}^{(D)}\equiv\sum^{3}_{j=2}P_{1j}^{(D)}\times\frac{1}{2}=\frac{\beta_{1}}{D_{11}}(1-P_{dd}^{(S)}),\qquad
P_{1d}^{\mathrm{(DP)}}\equiv\sum^{3}_{j=2}P_{1j}^{\mathrm{(DP)}}\times\frac{1}{2}=\frac{\gamma_{1}}{D_{11}}(1-P_{dd}^{(S)}).
\end{eqnarray}
Provided there is no dephasing noise at the donors ($\gamma_{d}=0$), the decay probability at the donors $P_{dd}^{(D)}=P_{dd}^{(S)}$ and it is larger than $1/2$. Reciprocally, the decay probability at the receiver is less than $1/2$ and so is the transfer probability to the RC. This explains the transfer efficiency $\epsilon<1/2$ in equation~\eref{eq:seproof}. On the other hand, if introducing and increasing the dephasing noise at the donors ($\gamma_{d}>0$), the decay probability at the donors decreases to zero, $P_{dd}^{(D)}=(\beta_{d}/D_{dd})P_{dd}^{(S)}\rightarrow 0$, whereas the dephasing probability at the donors $P_{dd}^{\mathrm{(DP)}}$ increases. Then, as seen in equation~\eref{eq:efficiency_expansion_1} or \eref{eq:efficiency_expansion_2}, the stochastic processes are summed up to result in the high transfer efficiency.

\begin{figure} [t]
\centering
\subfigure{\includegraphics{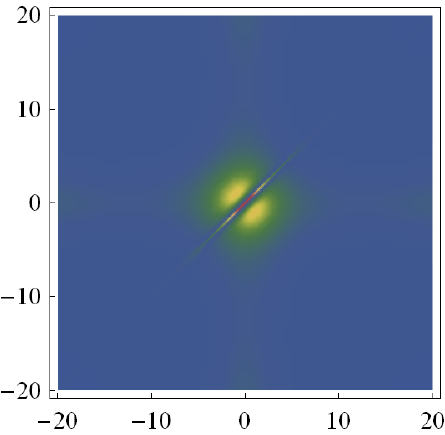}}\hspace{5pt}
\subfigure{\includegraphics[width=0.276\textwidth]{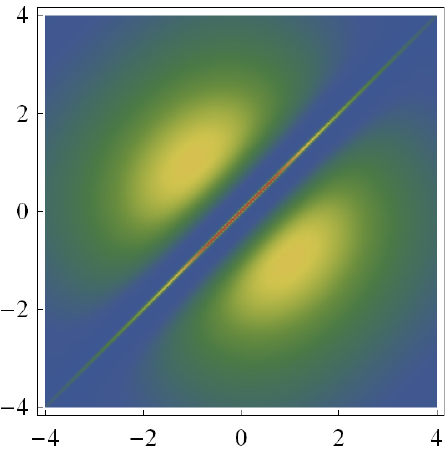}}\hspace{10pt}
\subfigure{\includegraphics{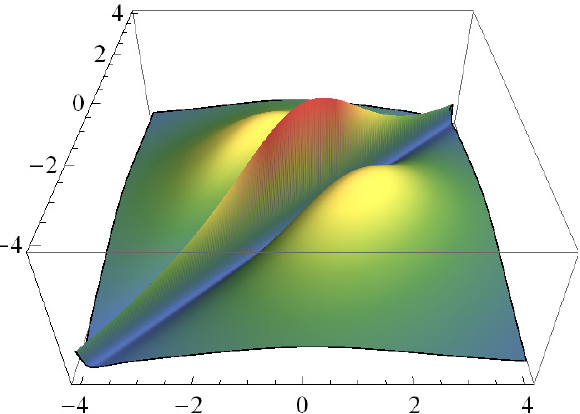}}\\
{\footnotesize (a) No dephasing noise}\\
\subfigure{\includegraphics{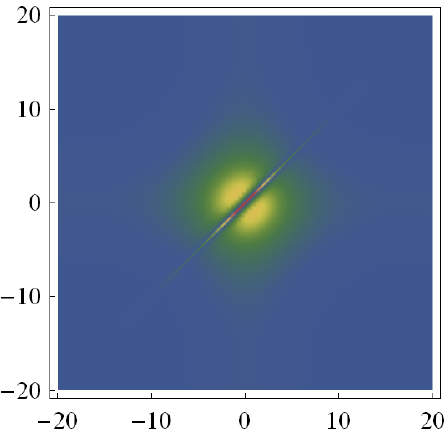}}\hspace{5pt}
\subfigure{\includegraphics[width=0.276\textwidth]{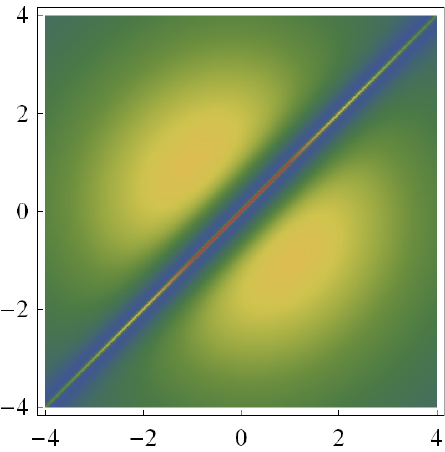}}\hspace{10pt}
\subfigure{\includegraphics{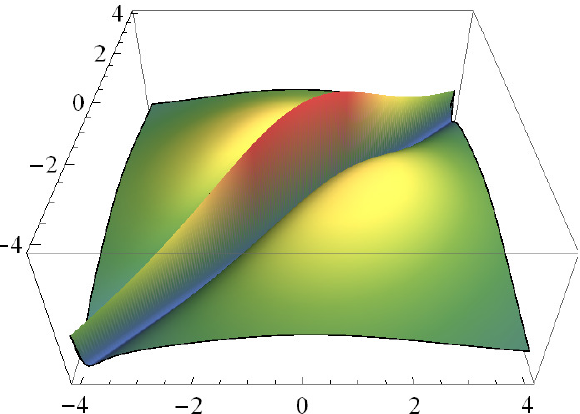}}\\
{\footnotesize (b) Dephasing at the receiver}\\
\subfigure{\includegraphics{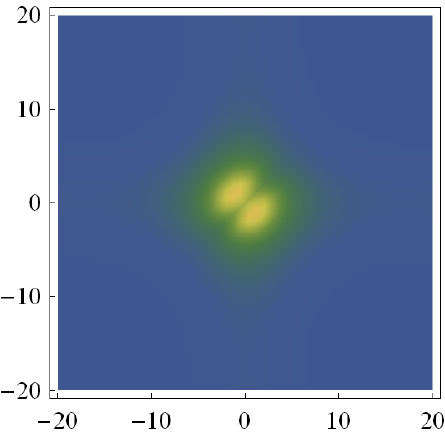}}\hspace{5pt}
\subfigure{\includegraphics[width=0.276\textwidth]{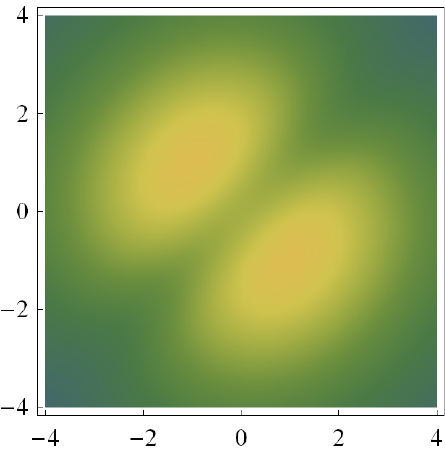}}\hspace{10pt}
\subfigure{\includegraphics{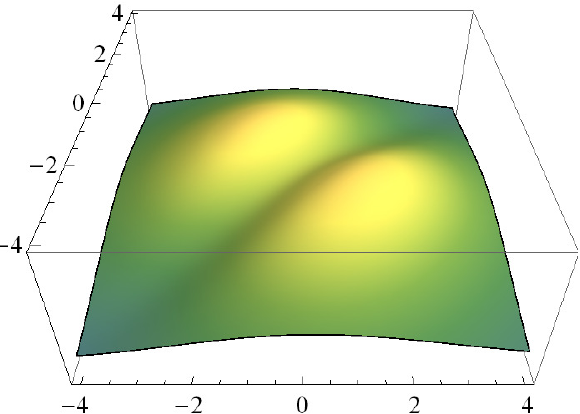}}\\
{\footnotesize (c) Dephasing at the donors}\\
\vspace{10pt}
\hspace{260pt}
\subfigure{\includegraphics[width=0.4\textwidth]{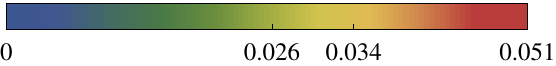}}
\caption{(color online) Inverse transfer time to the RC $(\tau_{\mathrm{RC}}J)^{-1}$ of the bi-pathway complex in \fref{fig:antenna_complexes}(b) with symmetric couplings constants $J=J_{12}=J_{13}$, parameterized by energy-level mismatches. In each panel, horizontal and vertical axes represent energy-level mismatches $\Delta\Omega_{12}/J$ and $\Delta\Omega_{13}/J$, respectively. The parameters are chosen as $\Gamma_{\mathrm{RC}}/J=10^{-1}$, $\Gamma_{j}/J=10^{-3}$ ($\beta_{j}=\Gamma_{j}+\Gamma_{\mathrm{RC}}\delta_{1j}$) and the dephasing constants $(\gamma_{1}/J,\gamma_{d}/J)$ = $(0,0)$, $(10^{-1},0)$, $(0,10^{-1})$ from the first to the third row.}
\label{fig:transfer_time_to_the_RC}
\end{figure}

\subsection{Temporal characteristics of Relaxation to the Ground State and of Transfer to the Reaction Center}

\begin{figure} [t]
\centering
\subfigure{\includegraphics{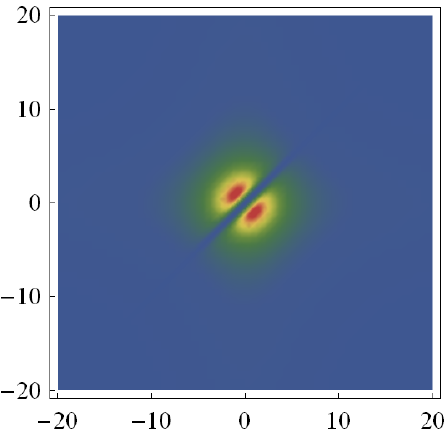}}\hspace{5pt}
\subfigure{\includegraphics[width=0.276\textwidth]{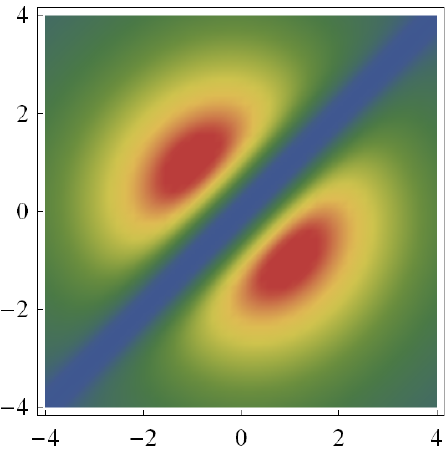}}\hspace{10pt}
\subfigure{\includegraphics{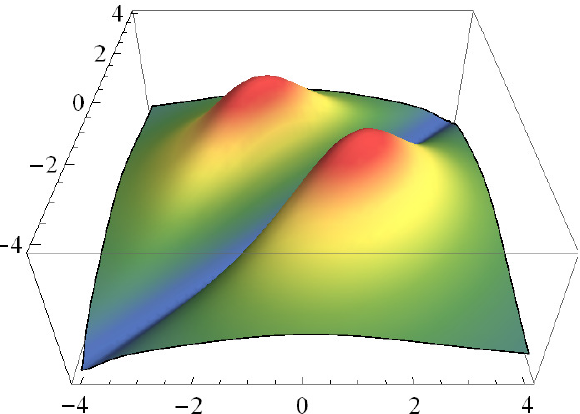}}\\
{\footnotesize (a) No dephasing noise}\\
\subfigure{\includegraphics{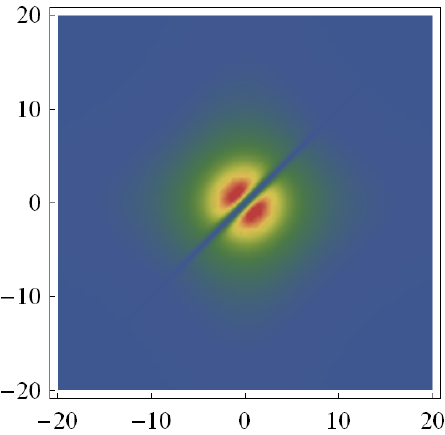}}\hspace{5pt}
\subfigure{\includegraphics[width=0.276\textwidth]{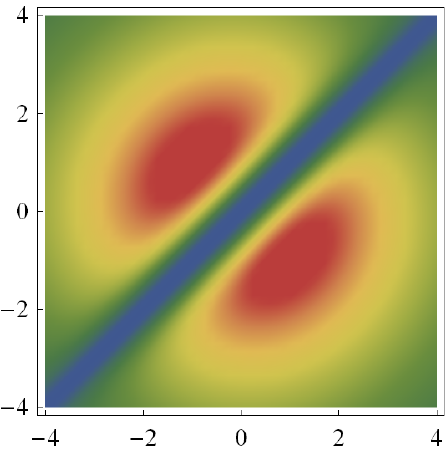}}\hspace{10pt}
\subfigure{\includegraphics{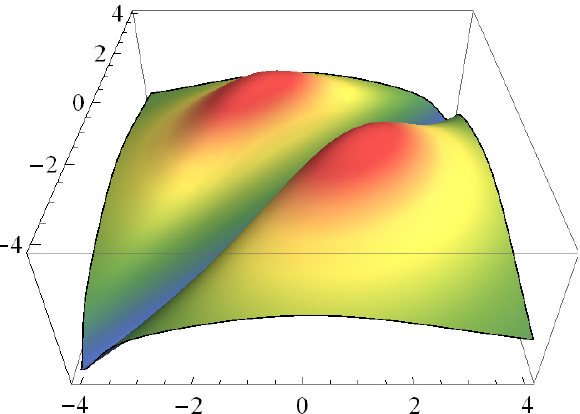}}\\
{\footnotesize (b) Dephasing at the receiver}\\
\subfigure{\includegraphics{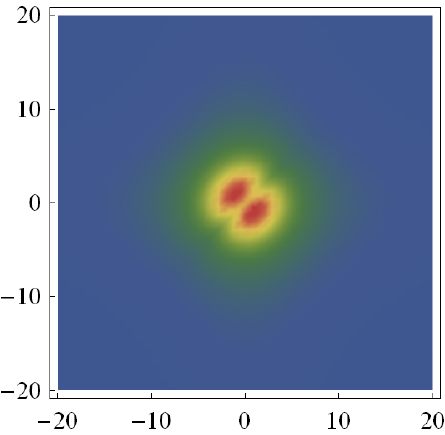}}\hspace{5pt}
\subfigure{\includegraphics[width=0.276\textwidth]{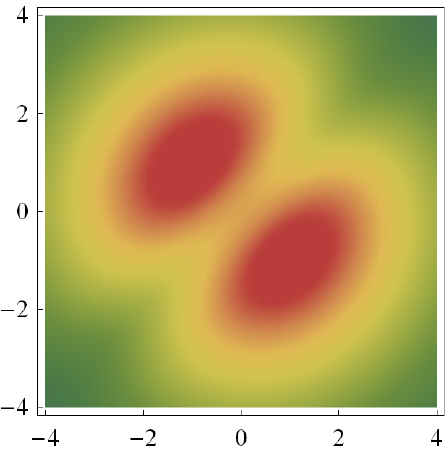}}\hspace{10pt}
\subfigure{\includegraphics{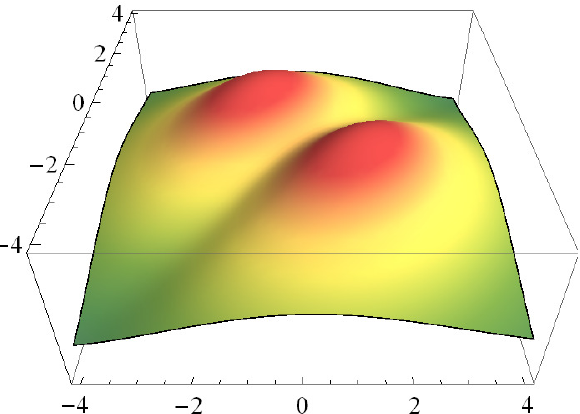}}\\
{\footnotesize (c) Dephasing at the donors}\\
\vspace{10pt}
\hspace{260pt}
\subfigure{\includegraphics[width=0.4\textwidth]{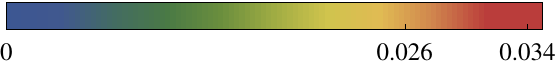}}
\caption{(color online) Inverse relaxation time $(\tau_{R}J)^{-1}$ of the bi-pathway complex in \fref{fig:antenna_complexes}(b) with symmetric couplings constants $J=J_{12}=J_{13}$, parameterized by energy-level mismatches. In each panel, horizontal and vertical axes represent energy-level mismatches $\Delta\Omega_{12}/J$ and $\Delta\Omega_{13}/J$, respectively. The parameters are chosen as $\Gamma_{\mathrm{RC}}/J=10^{-1}$, $\Gamma_{j}/J=10^{-3}$ ($\beta_{j}=\Gamma_{j}+\Gamma_{\mathrm{RC}}\delta_{1j}$) and the dephasing constants $(\gamma_{1}/J,\gamma_{d}/J)$ = $(0,0)$, $(10^{-1},0)$, $(0,10^{-1})$ from the first to the third row.}
\label{fig:relaxation_time}
\end{figure}

\begin{figure} [t]
\subfigure[Transfer Efficiency]{\includegraphics[width=0.41\textwidth]{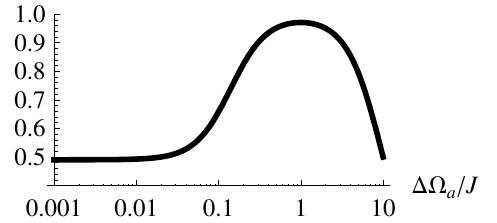}}\hspace{3pt}
\subfigure[Temporal Characteristics]{\includegraphics[width=0.41\textwidth]{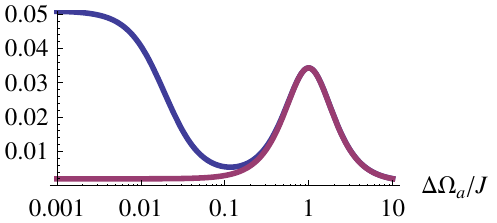}}
\subfigure{\includegraphics[width=0.16\textwidth]{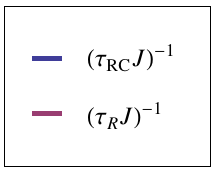}}
\caption{(color online) Transfer efficiency and temporal characteristics of the bi-pathway complex used in \fref{fig:transfer_time_to_the_RC} and \fref{fig:relaxation_time} for anti-symmetric energy-level mismatches $\Delta\Omega_{12}/J=-\Delta\Omega_{13}/J=\Delta\Omega_{a}/J$ with no dephasing noise. As the asymmetric degree of energy-level mismatches $\Delta\Omega_{a}/J$ is increased up to 0.1, (a) the transfer efficiency $\epsilon$ increases slightly higher than $1/2$ but (b) the inverse transfer time to the RC $(\tau_{\mathrm{RC}}J)^{-1}$ decreases close to the inverse relaxation time $(\tau_{R}J)^{-1}$. For $\Delta\Omega_{a}/J>0.1$, the difference between two temporal characteristics becomes negligible as the relaxation is mainly caused by the transfer to the RC.}
\label{fig:anti_symmetric_mismatches}
\end{figure}

\begin{figure} [t]
\subfigure[Transfer Efficiency]{\includegraphics[width=0.41\textwidth]{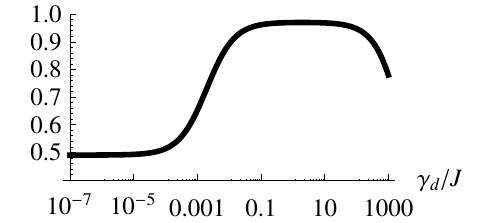}}\hspace{3pt}
\subfigure[Temporal Characteristics]{\includegraphics[width=0.41\textwidth]{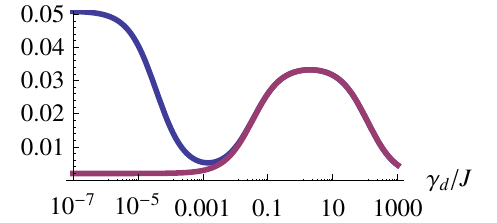}}
\subfigure{\includegraphics[width=0.16\textwidth]{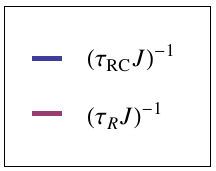}}
\caption{(color online) Transfer efficiency and temporal characteristics of the bi-pathway complex used in \fref{fig:transfer_time_to_the_RC} and \fref{fig:relaxation_time} for no energy-level mismatches $\Delta\Omega_{12}=\Delta\Omega_{13}=0$ with dephasing noise $\gamma_{d}$ at the donors. No dephasing noise at the receiver is assumed, $\gamma_{1}=0$.}
\label{fig:time_dephasing}
\end{figure}

\begin{table}
\caption{\label{table:time_characteristics}Transfer efficiency and temporal characteristics of the bi-pathway complex used in \fref{fig:transfer_time_to_the_RC} and \fref{fig:relaxation_time} for no energy-level mismatches $\Delta\Omega_{12}/J=\Delta\Omega_{13}/J=0$ and anti-symmetric energy-level mismatches $\Delta\Omega_{12}/J=-\Delta\Omega_{13}/J=\pm 1$.}
\begin{tabular*}{\textwidth}{@{}l*{15}{@{\extracolsep{0pt plus12pt}}l}}
\br
$\Delta\Omega_{12}/J$&$\Delta\Omega_{13}/J$&$\gamma_{1}/J$&$\gamma_{d}/J$&$(\tau_{\mathrm{RC}}J)^{-1}$&$(\tau_{R}J)^{-1}$&$\epsilon$\\
\mr
0&0&0&0&0.051&0.002&0.490\\
0&0&$10^{-1}$&0&0.051&0.002&0.490\\
0&0&0&$10^{-1}$&0.026&0.026&0.962\\
\mr
$\pm 1$&$\mp 1$&0&0&0.034&0.034&0.971\\
$\pm 1$&$\mp 1$&$10^{-1}$&0&0.034&0.034&0.971\\
$\pm 1$&$\mp 1$&0&$10^{-1}$&0.034&0.034&0.971\\
\br
\end{tabular*}
\end{table}

We shall investigate temporal characteristics of the single-exciton transfer. In order to consider relaxation process that the initially excited state decays to the ground state, we define relaxation time $\tau_{R}$ as
\begin{eqnarray}
\tau_{R}=\int^{\infty}_{0}dt\sum_{j=1}^{n}\beta_{j}\bra{j}\rho(t)\ket{j}t.
\end{eqnarray}
Here, $\sum_{j=1}^{n}\beta_{j}\bra{j}\rho(t)\ket{j}$ is the probability density function of time $t$ that the system state $\rho(t)$ decays to the ground state at $t$, satisfying the normalization condition, $\int^{\infty}_{0}dt\sum_{j=1}^{n}\beta_{j}\bra{j}\rho(t)\ket{j}=1$. If the probability of finding an exciton in the system is an exponentially decaying function $\exp(-t/\tau)$, $\tau=\tau_{R}$. Another relevant temporal characteristic is the transfer time to the RC, defined by
\begin{eqnarray}
\tau_{\mathrm{RC}}=\frac{1}{\epsilon}\int^{\infty}_{0}dt\Gamma_{\mathrm{RC}}\bra{1}\rho(t)\ket{1}t,
\end{eqnarray}
where $(\Gamma_{\mathrm{RC}}/\epsilon)\bra{1}\rho(t)\ket{1}$ is the probability density function of time $t$ that an exciton is transferred to the RC at $t$. When the time dependence of the transfer to the RC is given by a function $\epsilon(1-\exp(-t/\tau))$, $\tau=\tau_{\mathrm{RC}}$.

For a bi-pathway complex with couplings $J=J_{12}=J_{13}$, the inverse transfer time to the RC $(\tau_{\mathrm{RC}}J)^{-1}$ and relaxation time $(\tau_{R}J)^{-1}$, normalized by $J$, are presented in \fref{fig:transfer_time_to_the_RC} and \fref{fig:relaxation_time}, respectively. Note that the temporal characteristics have peaks at the points where the transfer efficiency $\epsilon$ is high in \fref{fig:energy_mismatches}(a)--(c). In the first row of panels in \fref{fig:transfer_time_to_the_RC} where no dephasing is assumed, the inverse transfer time is maximal at no energy-level mismatches and has a sharp peak along the symmetric energy-level mismatches $\Delta\Omega=\Delta\Omega_{12}=\Delta\Omega_{13}$. This is due to the fast transfer to the RC of the non-destructive components of $\ket{v_{p=2,3}}$ in equations~\eref{eq:pure_destructive_interference}--\eref{eq:pure_destructive_interference_2} (see \fref{fig:population_dynamics}). The fast transfer occurs in a probability less than $1/2$, while the destructive component of $\ket{v_{1}}$ results in the exciton-loss at the donors in a probability of $1/2$. The probability amplitudes of $\ket{v_{p=2,3}}$ decay with rate constants $\frac{1}{4}(\beta_{1}+\beta_{d})+(-1)^{p=2,3}\mathrm{Re}[Y]$ where $\mathrm{Re}[Y]$ is the real part of $Y$ in equation~\eref{eq:evalue23}. The inverse transfer time to the RC is determined by the probability of an exciton being at the receiver and for the symmetric energy-level mismatches it is upper bounded as
\begin{eqnarray}
\fl \tau_{\mathrm{RC}}^{-1}=\frac{\beta_{1}+\beta_{d}}{2}\left\{1+\frac{1}{2}\frac{(\beta_{1}+\beta_{d})^{4}+4(\beta_{1}-\beta_{d})^{2}\Delta\Omega^{2}}{[4(J_{12}^{2}+J_{13}^{2})+\beta_{1}\beta_{d}](\beta_{1}+\beta_{d})^{2}+4\beta_{1}\beta_{d}\Delta\Omega^{2}}\right\}^{-1}\le\frac{\beta_{1}+\beta_{d}}{2}.
\end{eqnarray}
The inverse transfer time approaches the upper bound if $\beta_{1},\beta_{d}\ll J_{12},J_{13}$ and $\Delta\Omega=0$, as in the case of the bi-pathway complex in \fref{fig:transfer_time_to_the_RC}: $(\tau_{\mathrm{RC}}J)^{-1}\approx\frac{1}{2}(\beta_{1}+\beta_{d})/J=0.051$. On the other hand, the probability amplitude of $\ket{v_{1}}$ decays with a rate constant $\frac{1}{2}\beta_{d}$ (see equation~\eref{eq:evalue1}). This causes the slow relaxation of the symmetric energy-level mismatches (see the first row of panels in \fref{fig:relaxation_time}):
\begin{eqnarray}
\fl \tau_{R}^{-1}=2\beta_{d}\left\{1+\frac{8(J_{12}^{2}+J_{13}^{2})(\beta_{1}+\beta_{d})\beta_{d}+\beta_{1}\beta_{d}(\beta_{1}+\beta_{d})^{2}+4\beta_{1}\beta_{d}\Delta\Omega^{2}}{[4(J_{12}^{2}+J_{13}^{2})+\beta_{1}\beta_{d}](\beta_{1}+\beta_{d})^{2}+4\beta_{1}\beta_{d}\Delta\Omega^{2}}\right\}^{-1}\le 2\beta_{d}.
\end{eqnarray}
For slightly-asymmetric energy-level mismatches, partially destructive interference occurs at the receiver molecule, which increases the transfer efficiency $\epsilon$ slightly higher than $1/2$ but decreases the inverse transfer time $(\tau_{\mathrm{RC}}J)^{-1}$ instead (see \fref{fig:anti_symmetric_mismatches} and table~\ref{table:time_characteristics}). As the asymmetric degree of energy-level mismatches increases, the interference effect eventually disappears and the temporal characteristics are optimized at the anti-symmetric energy-level mismatches $\Delta\Omega_{12}/J=-\Delta\Omega_{13}/J\approx\pm 1$. As in the third rows of panels in \fref{fig:transfer_time_to_the_RC} and \fref{fig:relaxation_time}, the interference effect also disappears as increasing the dephasing $\gamma_{d}$ at the donors (see \fref{fig:time_dephasing} and table~\ref{table:time_characteristics}).

\section{Multi-Pathway Complex}

\begin{figure} [t]
\centering
\subfigure[FMO complex]{\includegraphics{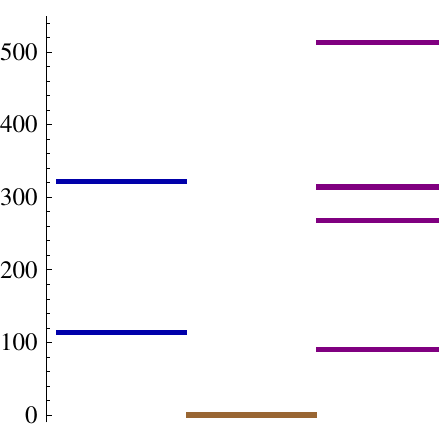}}\hspace{10pt}
\subfigure[$\gamma=0$]{\includegraphics{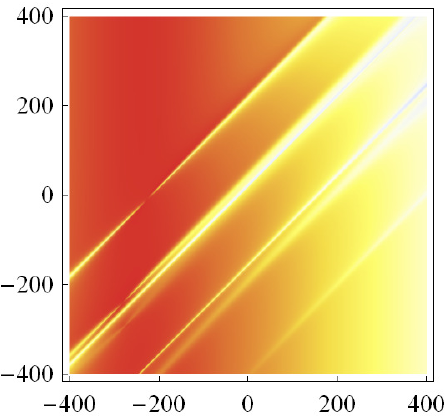}}\hspace{10pt}
\subfigure[$\gamma=10^{2}/188$]{\includegraphics{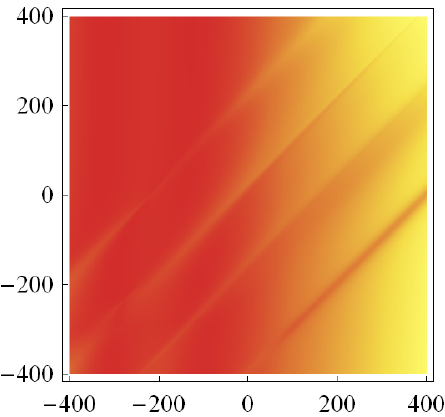}}\\
\vspace{10pt}
\hspace{235pt}
\subfigure{\includegraphics[width=0.4\textwidth]{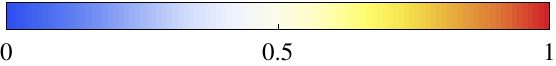}}
\caption{(color online) (a) Energy eigenvalues of two sub-complexes of the FMO complex (blue for one sub-complex with molecules 1 and 2 and purple for the other with molecules 4 to 7) and the excited energy of the receiver (brown). Here, we shifted the zero of energy by 12230 (all numbers are given in the units
of $\mathrm{cm}^{-1}$). Transfer efficiency of the energy-shifted FMO complex with (b) no dephasing noise and (c) uniform dephasing constants $\gamma_{j}=\gamma, \forall j$. In each panel, horizontal and vertical axes represent the energy shifts $\Delta\Omega_{s}$ of the sub-complexes $s=1,2$, respectively, and the parameters are chosen as $\Gamma_{\mathrm{RC}}=10^{3}/188$ and $\Gamma_{j}=1/188$, $\forall j$ ($\beta_{j}=\Gamma_{j}+\Gamma_{\mathrm{RC}}\delta_{1j}$).}
\label{fig:FMO_complex}
\end{figure}

We now briefly discuss the generalization to a multi-pathway complex consisting of sub-complexes with multi-level antenna molecules. Suppose that a receiver is coupled to independent sub-complexes.  When decoupled, each sub-complex yields a set of energy eigenvalues. It is found that {\it if any pair of sub-complexes share a common energy eigenvalue, there arises destructive interference at the shared molecule, which funnels the excitation energies from the sub-complexes.} In particular, consider the FMO complex consisting of seven BChl molecules, approximated as a bi-pathway complex where a receiver (molecule 3) is coupled to two sub-complexes, one ($s=1$) consisting of molecules 1 and 2 and the other ($s=2$) of molecules 4 to 7 (Here we use the usual numbering of the BChls, which was chosen by Fenna and Matthews~\cite{Fenna1975}). We present the energy eigenvalues of each sub-complex in \fref{fig:FMO_complex}(a), based on the Hamiltonian of a FMO monomer of {\it Prosthecochloris aestuarii} in Ref.~\cite{Adolphs2006},
\begin{equation}
H=
\begin{pmatrix}
215 & -104.1 & 5.1 & -4.3 & 4.7 & -15.1 & -7.8 \\
-104.1 & 220 & 32.6 & 7.1 & 5.4 & 8.3 & 0.8 \\
5.1 & 32.6 & 0 & -46.8 & 1 & -8.1 & 5.1 \\
-4.3 & 7.1 & -46.8 & 125 & -70.7 & -14.7 & -61.5 \\
4.7 & 5.4 & 1 & -70.7 & 450 & 89.7 & -2.5 \\
-15.1 & 8.3 & -8.1 & -14.7 & 89.7 & 330 & 32.7 \\
-7.8 & 0.8 & 5.1 & -61.5 & -2.5 & 32.7 & 280
\end{pmatrix},
\end{equation}
where we shifted the zero of energy by 12230 and all numbers are given in the units of $\mathrm{cm}^{-1}$. As shifting the energies of the two sub-complexes by $\Delta\Omega_{s=1,2}$, we present the transfer efficiency without/with dephasing in \fref{fig:FMO_complex}(b)/(c). Here, we assume that an exciton is created at molecule 1. It is clearly shown by seemingly four diagonal lines in \fref{fig:FMO_complex}(b) that destructive interference at the receiver results in the suppression of the transfer efficiency. The actual suppressing lines are eight but they are overlapped due to the almost equal energy-eigenvalue differences of the sub-complexes. In \fref{fig:FMO_complex}(c), it is shown that dephasing noise reduces the destructive interference, which consequently enhances the transfer efficiency of the FMO complex ($\Delta\Omega_{s}=0$) from about $77{\%}$ to $96{\%}$ when $\gamma=10^{2}/188$. We would note that there can arise destructive interference inside a sub-complex if it has a multi-pathway structure depending on the excited energies and electronic coupling constants of the constituent molecules.

\section{Remark}

In summary, we investigated the role of energy-level mismatches between sub-complexes in a multi-pathway complex of a light-harvesting complex. For a single-pathway complex, we showed that the presence of energy-level mismatches never enhances the transfer efficiency, as it is likely to cause the Anderson localization that reduces the energy transfer through the energy pathway. However, we demonstrated that for a multi-pathway complex, the absence of energy-level mismatches causes the quantum destructive interference at the shared molecule, which funnels the excitation energies from the sub-complexes. We showed that the undesired destructive interference diminishes as introducing energy-level mismatches so that the transfer efficiency is significantly improved, even though they may cause the Anderson localization effect and suppress the energy transfer in each energy pathway. Due to the competition between the localization effect and the destructive interference at the shared molecule, it is crucial to find an optimal amount of energy-level mismatches so as to maximize the transfer efficiency as minimizing both the undesired effects. We also showed that the quantum dephasing noise relaxes the energy localization and destroys the destructive interference at the shared molecule, implying that the cooperation of the energy-level mismatches and the dephasing noise significantly increases the transfer efficiency after all. This mechanism will be qualitatively unaltered even if considering non-Markovian noise~\cite{Ishizaki2009,Ishizaki2010}, as long as the role of the noise is to destroy the destructive interference at the shared molecule. The results presented here suggest that the energy-level mismatches typically given in natural photosynthetic complexes can play a crucial role in the efficient energy transfer. It is a timely question to investigate the optimal energy landscape with energy-level mismatches and non-uniform electronic couplings in a more elaborate dephasing model which accounts the thermal fluctuations of surrounding proteins and the intra/inter-molecular vibrations~\cite{Renger2001}.

\ack
We are very grateful to M. B. Plenio for useful comments. This CRI work was supported by the National Research Foundation of Korea (NRF) grant funded by the Korea government (MEST) (No. 3348-20100018 and No. 2010-0015059).


\section*{References}
\begin{thebibliography}{}

\bibitem{Engel2007} G. S. Engel, T. R. Calhoun, E. L. Read, T-K. Ahn, T. Mancal, Y-C. Cheng, R. E. Blackenship and G. R. Fleming 2007 {\it Nature} {\bf 446} 782

\bibitem{Lee2007} H. Lee, Y-C. Cheng, G. R. Fleming 2007 {\it Science} {\bf 316} 1462

\bibitem{Collini2010} E. Collini, C. Y. Wong, K. E. Wilk, P. M. G. Curmi, P. Brumer and G. D. Scholes 2010 {\it Nature} {\bf 463} 644

\bibitem{Mohseni2008} M. Mohseni, P. Rebentrost, S. Lloyd and A. Aspuru-Guzik 2008 {\it J. Chem. Phys.} {\bf 129} 174106

\bibitem{Plenio2008} M. B. Plenio and S. F. Huelga 2008 {\it New J. Phys.} {\bf 10} 113019

\bibitem{Caruso2009} F. Caruso, A. W. Chin, A. Datta, S. F. Huelga, and M. B. Plenio 2009 {\it J. Chem. Phys.} {\bf 131} 105106

\bibitem{Chin2010} A. W. Chin, A. Datta, F. Caruso, S. F. Huelga and M. B. Plenio 2010 {\it New J. Phys.} {\bf 12} 065002

\bibitem{Ishizaki2009_2} A. Ishizaki and G. R. Fleming 2009 {\it J. Chem. Phys.} {\bf 130} 234111

\bibitem{Ishizaki2009} A. Ishizaki and G. R. Fleming 2009 {\it PNAS} {\bf 106} 17255

\bibitem{Panitchayangkoon2010} G. Panitchayangkoon, D. Hayes, K. A. Fransted, J. R. Caram, E. Harel, J. Wen, R. E. Blankenship, and G. S. Engel 2010 {\it PNAS} {\bf 107} 12766

\bibitem{Cheng2006} Y. C. Cheng and R. J. Silbey 2006 {\it Phys. Rev. Lett.} {\bf 96} 028103

\bibitem{Nazir2009} A. Nazir 2009 {\it Phys. Rev. Lett.} {\bf 103} 146404

\bibitem{Calhoun2009} T. R. Calhoun, N. S. Ginsberg, G. S. Schlau-Cohen, Y-C. Cheng, M. Ballottari, R. Bassi and G. R. Fleming 2009 {\it J. Phys. Chem. B} {\bf 113} 16291

\bibitem{Rebentrost2009} P. Rebentrost, M. Mohseni, I. Kassal, S. Lloyd and A. Aspuru-Guzik 2009 {\it New J. Phys.} {\bf 11} 033003

\bibitem{Rebentrost2009} P. Rebentrost, M. Mohseni and A. Aspuru-Guzik 2009 {\it J. Phys. Chem. B} {\bf 113} 9942

\bibitem{Anderson1958} P. W. Anderson 1958 {\it Phys. Rev.} {\bf 109}, 1942

\bibitem{Cheng2009} Y-C. Cheng and G. R. Fleming 2009 {\it Annu. Rev. Phys. Chem.} {\bf 60} 241

\bibitem{Ishizaki2010} A. Ishizaki, T. R. Calhoun, G. S. Schlau-Cohen and G. R. Fleming 2010 {\it Phys. Chem. Chem. Phys.} {\bf 12} 7319

\bibitem{Gardiner1991} C. W. Gardiner and P. Zoller 1991 {\it Quantum Noise} (Springer-Verlag, Berlin)

\bibitem{Breuer2002} H-P. Breuer and F. Petruccione 2002 {\it The Theory of Open Quantum Systems} (Oxford University Press, Oxford)

\bibitem{Fenna1975} R. E. Fenna and B. W. Matthews 1975 {\it Nature} {\bf 258} 573

\bibitem{Adolphs2006} J. Adolphs and T. Renger 2006 {\it Biophys. J.} {\bf 91} 2778

\bibitem{Renger2001} T. Renger, V. May and O. K\"{u}hn 2001 {\it Phys. Rep.} {\bf 343} 137

\end{thebibliography}
\end{document}